\documentclass[twocolumn,10pt,prl,aps,showpacs]{revtex4}
\def \D {\mbox{D}}
\def \D {\mbox{D}}

\def \curl {\mbox{curl}\,}

\def\3nab{\tilde{\nabla}}

\def\hs {\,-\,}

\def\be {\begin{equation}}
\def\ee {\end{equation}}
\def\bea {\begin{eqnarray}}
\def\eea {\end{eqnarray}}

\newcommand{\sfrac}[2]{{\textstyle{#1\over#2}}}
\def\case#1/#2{\textstyle\frac{#1}{#2}}
\usepackage{graphicx}% Include figure files
\usepackage{dcolumn}% Align table columns on decimal point
\usepackage{bm}% bold math
\usepackage{longtable}

\begin{document}

%%%%%%%%%%%%%%%%%%%%%%%%%%%%%%%%%%%%%%%%%%%%%%%%%%%%
\title{Are braneworlds born isotropic?}
%%%%%%%%%%%%%%%%%%%%%%%%%%%%%%%%%%%%%%%%%%%%%%%%%%%%

\author{ Peter K. S. Dunsby${}^{1,2}$, Naureen Goheer${}^1$, 
Marco Bruni${}^{3}$ and Alan Coley${}^{4}$}

\affiliation{1. Department of Mathematics and Applied Mathematics,
  University of Cape Town, 7701 Rondebosch, Cape Town, South Africa}

\affiliation{2. South African Astronomical Observatory, 
  Observatory 7925, Cape Town, South Africa.}

\affiliation{3. Institute of Cosmology and Gravitation, University of
Portsmouth, Mercantile House, Portsmouth PO1 2EG}

\affiliation{4. Department of Mathematics and  Statistics, Dalhousie 
University, Halifax, Nova Scotia, Canada B3H 3JS}

\date{December 16, 2003}

%%%%%%%%%%%%%%%%%%%%%%%%%%%%%%%%%%%%%%%%%%%%%%%%%%%%% 
\begin{abstract}
It has recently been suggested that an isotropic singularity may be 
a generic feature of brane cosmologies, even in the inhomogeneous 
case. Using the covariant and gauge\hs invariant approach we present 
a detailed analysis of linear perturbations of the isotropic model 
${\cal F}_b$ which is a past attractor in the phase space of homogeneous 
Bianchi models on the brane. We find that for matter with an equation 
of state parameter $\gamma > 1$, the dimensionless variables representing 
generic anisotropic and inhomogeneous perturbations decay as $t\rightarrow 0$, 
showing that the model ${\cal F}_b$ is asymptotically stable in the past. 
We conclude that brane universes are born with isotropy naturally 
built\hs in, contrary to standard cosmology. The observed large\hs scale 
homogeneity and isotropy of the universe can therefore be explained 
as a consequence of the initial conditions if the brane\hs world paradigm 
represents a description of the very early universe. 
\end{abstract}

%%%%%%%%%%%%%%%%%%%%%%%%%%%%%%%%%%%%%%%%%%%%%%%%%%%%%%%
\maketitle
%%%%%%%%%%%%%%%%%%%%%%%%%%%%%%%%%%%%%%%%%%%%%%%%%%%%%%%%%
\section{Introduction}
%%%%%%%%%%%%%%%%%%%%%%%%%%%%%%%%%%%%%%%%%%%%%%%%%%%%%%%%%
A classic problem of cosmology is finding ways to explain the very 
high degree of isotropy observed in the Cosmic Microwave Background (CMB). 
In general relativity, where isotropy is a special rather than generic 
feature of cosmological models, we need a dynamical mechanism which is 
able to produce isotropy. One of the most efficient mechanisms for 
isotropising the universe is Inflation \footnote{There are perturbative proofs
of the cosmic no-hair conjecture, i.e.\ classical perturbations in
inflationary models with a scalar field or cosmological constant
are swept out, as well as (local) proofs for homogeneous and
inhomogeneous models (see e.g.~\cite{BMT} and references
therein).} but it requires sufficiently homogeneous initial data in order
to start at all~\cite{KT}. Although one can adopt the view that it 
is sufficient to have one such homogeneous enough patch in an otherwise 
inhomogeneous initial universe to explain what we observe, this
seems somehow unsatisfactory. In our view the isotropy problem 
remains open to debate in standard cosmology.

Over the past few years the brane-world paradigm has received 
considerable attention as a possible candidate for string inspired 
cosmology (see~\cite{roy3} for a recent review). In this scenario 
the observable universe is a 4\hs dimensional slice, {\it the brane}, 
embedded in a higher dimensional spacetime called {\it the bulk}.
Here we consider the formulation developed in~\cite{SMS} 
in order to generalise a previous model by Randall and Sundrum 
\cite{RS}, where the bulk is 5\hs dimensional and contains only 
a cosmological constant, assumed to be negative.  

In a series of recent papers a number of authors
~\cite{SC2,SVF,SC1,coley2,coley1,MS,HCY,HI,naureen1,naureen2} have presented 
a detailed description of the dynamics of homogeneous and anisotropic 
brane-worlds, finding a remarkable result: unlike standard general 
relativity, where the generic cosmological singularity is anisotropic, 
the past attractor for homogeneous anisotropic 
models in the brane is a simple Robertson\hs Walker model 
${\cal F}_b$. More significantly, in~\cite{coley2,coley1} it was 
shown that this result holds true in Bianchi IX models, as well as 
for some simple inhomogeneous models. Since the Belinski\hs Lifshitz\hs 
Kalatnikov (BLK) conjecture \cite{BLK} suggests that the 
Bianchi IX behaviour in the vicinity of the singularity is 
general, i.e. that the approach to the cosmological singularity 
in a generic inhomogeneous universe model should locally be the 
same as in Bianchi IX, it has been suggested that the isotropic 
singularity could be a generic feature of brane cosmological models.

If this conjecture \cite{coley2,coley1} could be proved correct,
brane cosmology would have the very attractive feature of having
isotropy built in. Inflation in this context would still be the
most efficient way of producing the fluctuations seen in the CMB,
but there would be no need of special initial conditions for it to
start \cite{GW}. Also the Penrose conjecture~\cite{penrose} on gravitational 
entropy and an initially vanishing Weyl tensor would be satisfied in 
these models (c.f.~\cite{tod,GCW}).

In this paper we prove that this conjecture is true, within a perturbative
approach and in the large\hs scale and high energy regime, as justified below.
We arrive at this result through a detailed analysis of generic linear 
inhomogeneous and anisotropic perturbations \cite{EB,BDE,goode,DBE} 
of this past attractor ${\cal F}_b$. This is done by using the full set 
of linear 1+3 covariant {\it propagation} and {\it constraint} equations 
for this background, which describe the kinematics of the fluid flow and 
the dynamics of the gravitational field (see eqns. 87-100 in \cite{roy2}). 
These equations are then split into their {\it scalar}, {\it vector} 
and {\it tensor} contributions giving three sets of linear propagation 
and constraint equations which govern the complete perturbation dynamics.  

From a dynamical systems point of view the past attractor ${\cal F}_b$ for 
brane homogeneous cosmological models found in~\cite{coley2,coley1} is a 
fixed point in the phase space of these models. This phase space may be 
thought of as an invariant submanifold  within an higher dimensional phase 
space for more general inhomogeneous  models. Our analysis can be 
thought of as an exploration of the neighbourhood of ${\cal F}_b$ out of 
the invariant submanifold explored in~\cite{coley2,coley1}. 

We restrict our analysis to large\hs scales, at a time when 
physical scales of perturbations are much larger than the Hubble 
radius, $\lambda\gg H^{-1}$ (equivalent to neglecting Laplacian 
terms in the evolution equations). This may at first glance seem 
restrictive, but it is not the case for the non\hs inflationary perfect 
fluid models that are relevant to our discussion. Indeed, it is well known 
that any wavelength $\lambda<H^{-1}$ at a given time becomes much larger 
than $H^{-1}$ at early enough times. Because of this crucial property
of perturbations for non\hs inflationary models our analysis is
completely general, i.e.\ valid for any $\lambda$ as $t\rightarrow 0$.

In what follows we restrict our analysis to the case of vanishing 
background dark radiation (${\cal U}=0$).

\begin{table*}
\caption{\label{tab:table1} Large-scale behaviour or geometric and kinematic 
quantities in the high energy limit. Here $p=3\gamma-3$, $q=6\gamma-2$, $r=-3$, $s=6\gamma-4$, $u=6\gamma-5$, $v=6\gamma-3$}
\begin{ruledtabular}
\begin{tabular}{cccc}
Quantity  & Scalar contribution & Vector Contribution & Tensor 
Contribution \\ \hline
$\Sigma$       &  $-\case{3}/{2}Q^*_0a^p\;,
                   -\case{6\gamma+1}/{\gamma(3\gamma+1)}
                  \Delta_1a^q$ & $-2Q^*_0a^p\;, 
               -\case{4}/{3\gamma}\case{1+6\gamma}/{1+3\gamma}\Delta_1a^q$ & 
                  $\Sigma_0a^p\;,\Sigma_1a^q$ \\
${\cal E}$     &  $-\case{3}/{2}Q_0^*a^p\;,\case{3(6\gamma+1)}/{3\gamma+1}
                  \Delta_1a^q$  & $-2Q^*_0a^p\;,
                  \case{4(1+6\gamma)}/{1+3\gamma}a^q$ & 
                  $\Sigma_0a^p\;,
               -3\gamma\Sigma_1a^q$   \\
$\Delta$       &  $\Delta_1a^q\;,\case{1}/{2}Q_1^*a^r\;,
                  -\case{\gamma}/{2(6\gamma-1)}U^*_0a^s$ & $\Delta_1a^q\;,
                  \case{1}/{2}Q^*_1a^r\;,
               -\case{1}/{2}\case{\gamma}/{6\gamma-1}U^*_0a^s$ & 0 \\
${Z}^*$   &  $-\case{3\gamma+1}/{\gamma}\Delta_1a^q\;, 
                  \case{3}/{2}Q^*_1a^r\;, 
                    \case{3\gamma-1}/{2(6\gamma-1)}U^*_0a^s$ 
               & $-\case{3\gamma+1}/{\gamma}\Delta_1a^q\;,\case{3}/{2}Q^*_1a^r\;, 
                 \case{3\gamma-1}/{2(6\gamma-1)}U^*_0a^s$ 
               & 0 \\
$Q^*$          &  $Q^*_0a^p\;,-\case{6\gamma}/{3\gamma+1}\Delta_1a^q\;,                             Q^*_1a^r\;,\case{3\gamma-1}/{3(6\gamma+1)}U^*_0a^s$ & 
                    $Q^*_0a^p\;,-\case{6\gamma}/{3\gamma+1}\Delta_1a^q\;,
                    Q^*_1a^r\;,
                    \case{1}/{3}\sfrac{3\gamma-1}{6\gamma-1}U^*_0a^s$ & 0 \\
$U^*$          &  $U^*_0a^s$ &  $U^*_0a^s$ & 0 \\
$W$          & 0 & $W_0a^u$ & 0 \\
${\cal H}$     & 0 & ${\cal H}_0a^s$ & ${\cal H}_0a^s\;,
               {\cal H}_1a^{\case{3}/{2}q}$ \\
$\bar{W}^*$    & 0 & 0 & 0 \\ 
$\bar{Q}^*$    & 0 & $-{\cal H}_0a^s\;, 12\gamma W^*_0a^u$ & 0 \\
$\bar{\Sigma}$ & 0 & ${\cal H}_0a^s$ &  ${\cal H}_0a^s\;,{\cal H}_1a^{\case{3}/{2}q}$ \\
$\bar{\cal E}$ & 0 & ${\cal H}_0a^s$ &  ${\cal H}_0a^s\;,-3\gamma{\cal H}_1a^{\case{3}/{2}q}$ \\
$\bar{\cal H}$ & 0 & 0 & 0 \\

\end{tabular}
\end{ruledtabular}
\end{table*}

%%%%%%%%%%%%%%%%%%%%%%%%%%%%%%%%%%%%%%%%%%%%%%%%%%%%%%%%%
\section{Dimensionless variables and harmonics}
%%%%%%%%%%%%%%%%%%%%%%%%%%%%%%%%%%%%%%%%%%%%%%%%%%%%%%%%%

We define dimensionless expansion normalised quantities for 
the shear $\sigma_{ab}$, the vorticity $\omega_a$, the 
electric $E_{ab}$ and magnetic $H_{ab}$ parts of the Weyl 
tensor \cite{goode}:
\begin{equation}
\Sigma_{ab}=\frac{\sigma_{ab}}{H}\;,~~
W_a=\frac{\omega_a}{H}\;,~~
{\cal E}_{ab}=\frac{E_{ab}}{H^2}\;,~~
{\cal H}_{ab}=\frac{H_{ab}}{H^2}\;,
\end{equation}
where $H$ is the Hubble parameter $H=\dot{a}/a$ and $a$ is the 
scale factor. 

It turns out to be convenient to use the 
dimensionless variable 
\begin{equation}
W^*_a = a\omega_a
\end{equation} 
to 
characterise the vorticity of the fluid flow. Here 
$\omega_a=\eta_{abc}\omega^{bc}$ is the usual 
vorticity vector.

The appropriate dimensionless density and expansion gradients 
which describe the {\it scalar} and {\it vector} parts of  
density perturbations are given by (see \cite{roy2} for details of definitions)
\bea {\Delta}_a=\frac{a}{\rho} \D_a \rho\;,~~ 
Z^*_a=\frac{3a}{H}\D_a H\;, \label{eq:grads}
\eea
and for the brane\hs world contributions we define the following
dimensionless gradients describing inhomogeneity in the non\hs 
local quantities
\be
U^*_a=\frac{\kappa^2\rho}{H^2}U_a\;,~~
Q^*_a=\frac{\kappa^2 a\rho}{H}Q_a\;,
\ee
where $U_a$ and $Q_a$ are defined in equation (27) of \cite{GM}. 

When undertaking a complete analysis of the perturbation dynamics it turns out
useful to define a set of auxiliary variables corresponding to the
curls of the standard quantities defined above:
\bea
\bar{W}^*_a&=&\frac{1}{H}\curl{W^*_a}\;,~~ \bar{\Sigma}_{ab}=\frac{1}{H}
\curl{\Sigma_{ab}}\;,   \\
\bar{{\cal E}}_{ab}&=&\frac{1}{H}\curl{\cal E}_{ab} \;,~~
\bar{\cal H}_{ab}=\frac{1}{H}{\cal H}_{ab}\;,\\
\bar{Q}^*_a&=&\frac{1}{H}\curl{Q}^*_a\;.
\eea
Finally, it is useful to use the dimensionless time derivative 
$d/d{\tau}=d/d~(ln~a)$ (denoted by a prime) to analyse the past 
evolution of the perturbation dynamics. 

We use the harmonics defined in~\cite{BDE} to expand the above 
tensors $X_a\;,X_{ab}$ in terms of scalar (S), vector (V) and tensor (T) 
harmonics $Q$. This yields a covariant and gauge invariant splitting 
of the evolution and constraint equations for the above quantities 
into three sets of evolution and constraint equations for scalar, vector 
and tensor modes respectively:
\bea
X&=&X^S Q^S \\
X_a&=&k^{-1}X^S Q_{a}^S+X^V Q_{a}^V \\
X_{ab}&=&k^{-2}X^S Q_{ab}^S+k^{-1}X^VQ_{ab}^V+X^T Q_{ab}^T\;.
\eea

%%%%%%%%%%%%%%%%%%%%%%%%%%%%%%%%%%%%%%%%%%%%%%%%%%%%%%%%%
\section{Perturbation Equations and their solutions}
%%%%%%%%%%%%%%%%%%%%%%%%%%%%%%%%%%%%%%%%%%%%%%%%%%%%%%%%%
We begin by giving the {\it evolution} equations 
for {\it scalar perturbations} (suppressing the label S):
\bea
\Sigma'&=&(3\gamma-2)\Sigma-{\cal E}\;,\\
 {\cal E}'&=&(6\gamma-3){\cal E}-3\gamma\Sigma\;,\\
{\Delta}'&=&(3\gamma-3)\Delta-\gamma{Z^*}\;,\\
{Z^*}'&=&(3\gamma-2){Z^*}-3(3\gamma+1)\Delta -{U^*}\;,\\
{Q^*}'&=&(3\gamma -3){Q^*}-\case{1}/{3}U^*-6\gamma\Delta\;,\\
{U^*}'&=&(6\gamma-4){U^*}\;.
\eea
The {\it scalar} constraint equations are:
\bea
2{\Sigma}&=&2Z^*-3{Q^*}\;,\label{eq:Sig}\\
2{\cal E}&=&6\Delta-3{Q^*}+{U^*} \label{eq:divE} \;.
\eea
{\it Vector perturbations} on large\hs scale are described by the 
following {\it evolution} equations for the basic variables
\bea
\Sigma'&=&(3\gamma-2)\Sigma-{\cal E}-2(\gamma-1)\bar{W}^*\;,\\
 {\cal E}'&=&(6\gamma-3){\cal E}-3\gamma\Sigma+6\gamma\bar{W}^*\;,\\
{\Delta}'&=&(3\gamma-3)\Delta-\gamma{Z^*}\;,\\
{Z^*}'&=&(3\gamma-2){Z^*}-(9\gamma+3)\Delta -{U^*}-(6\gamma-6)\bar{W}^*\;,\\
{Q^*}'&=&(3\gamma -3){Q^*}-\case{1}/{3}U^*-6\gamma\Delta\;,\\
{U^*}'&=&(6\gamma-4){U^*}\;,\\
{W^*}'&=&(3\gamma-4)W^*\;,\\
{\cal H}'&=&(6\gamma-3){\cal H}-12\gamma W^*+\bar{Q}^*\;,
\eea
and their curls
\bea
\bar{W}^{*'}&=&(6\gamma-5)\bar{W}^*\;,\\
\bar{Q}^{*'}&=&(6\gamma-4)\bar{Q}^*-36\gamma^2 W^*\;, \\
{\bar{\cal H}}'&=&(9\gamma-4)\bar{\cal H}-6\gamma\bar{W}^*\\
{\bar{\cal E}}'&=&(9\gamma-4)\bar{\cal E}-3\gamma\bar{\Sigma}\;,\\
{\bar{\Sigma}}'&=&(6\gamma-3)\bar{\Sigma}-\bar{\cal E}\;.
\eea
The {\it constraints} for the basic variables are
\bea
3\Sigma&=&4{Z^*}-6{Q^*}+6\bar{W}^*\;,\\
3{\cal E}&=&12\Delta-6{Q^*}+2{U^*}\;,\\
{\cal H}&=&12\gamma{W^*}-\bar{Q}^*\;,\\
\bar{\Sigma}&=&{\cal H}\;,\label{gm}
\eea
and the curl constraints are 
\bea
\bar{\cal H}&=&6\gamma\bar{W}^*\;,\\
\bar{\Sigma}&=&12\gamma{W^*}-\bar{Q}^*\;,\\
\bar{\cal E}&=&12\gamma{W^*}-\bar{Q}^*\;.
\eea

Finally, the large\hs scale evolution of {\it tensor perturbations} are 
governed by propagation equations for the tensor contributions 
of the shear $\Sigma_{ab}$, the electric ${\cal E}_{ab}$ and 
magnetic ${\cal H}_{ab}$ parts of the Weyl tensor:
\bea
{\Sigma}'&=&(3\gamma-2)\Sigma -{\cal E}\;,\\
{\cal E}'&=&(6\gamma-3){\cal E}-3\gamma\Sigma+\bar{\cal H}\;,\\
{\cal H}'&=&(6\gamma-3){\cal H}-\bar{\cal E}\;,
\eea
together with propagation equations for their curls
 \bea
 {\bar{\Sigma}}'&=&(6\gamma-3)\bar{\Sigma}-\bar{\cal E}\;,\\
 {\bar{\cal E}}'&=&(9\gamma-4)\bar{\cal E}-3\gamma\bar{\Sigma}\;,\\
 {\bar{\cal H}}'&=&(9\gamma-4)\bar{\cal H}\;.
 \eea
The only tensorial contribution to the constraints is
\be
\bar{\Sigma}={\cal H}\;.
\ee

Solutions to these three sets of equations are presented in Table I. On 
close inspection of the exponents we conclude that ${\cal F}_b$ is 
stable in the past (as $t\rightarrow 0$) to generic inhomogeneous 
and anisotropic perturbations provided the matter is described by 
a non\hs inflationary perfect fluid with $\gamma$\hs law equation of state 
parameter satisfying $\gamma > 1$ and we use the large\hs scale approximation 
($\lambda\gg H^{-1}$). In particular it can be easily seen that the 
expansion normalised shear vanishes as $a\rightarrow 0$, signalling 
isotropization.

%%%%%%%%%%%%%%%%%%%%%%%%%%%%%%%%%%%%%%%%%%%%%%%%%%%%%%%%%
\section{Discussion of results}
%%%%%%%%%%%%%%%%%%%%%%%%%%%%%%%%%%%%%%%%%%%%%%%%%%%%%%%%%

We have considered here only the case of vanishing background Weyl energy 
density, ${\cal U}=0$. This assumption considerably simplifies the analysis, 
but it is expected that our results will remain true for ${\cal U}\not =0$. 
Indeed, when ${\cal U}\not=0$ ${\cal F}_b$ still remains a past attractor 
of the isotropic models. In other words, our analysis is restricted to 
the invariant submanifold ${\cal U}=0$ of the larger phase space 
with ${\cal U}\not =0$, but this submanifold is asymptotically stable 
against ${\cal U}\not =0$ perturbations. A more complete analysis including 
this case will be the subject of a future investigation.

In a related paper the dynamics of a class of {\em spatially inhomogeneous 
$G_{2}$} cosmological models in the brane-world scenario has been studied 
\cite{CHL}. Area expansion normalized scale\hs invariant dependent variables,
the timelike area gauge and an effective logarithmic time $t$ were 
employed, and the initial singularity occurs for $t\rightarrow -\infty$. 
The resulting governing system of evolution equations of the spatially
inhomogeneous $G_{2}$ brane cosmological models can then be written
as a constrained system of autonomous first\hs order partial differential
equations in two independent variables.
The local dynamical behaviour of this class of spatially
inhomogeneous models close to the singularity was then studied
{\em numerically}. It was found that the area expansion
rate increases without bound (and hence the Hubble rate
$H \rightarrow \infty$ as $t \rightarrow -\infty$,
so that there always exists an initial singularity). For $\gamma > 4/3$, 
the numerics indicate isotropization towards ${\cal F}_b$ as
$t\rightarrow -\infty$ for {\em all} initial conditions. In the case 
of radiation ($\gamma=4/3$), the models were still found to isotropize 
as $t\rightarrow -\infty$, albeit slowly. From the numerical analysis 
we find that there is an initial isotropic singularity in all of 
these $G_2$ spatially inhomogeneous brane cosmologies for a range 
of parameter values which include the physically important cases of 
radiation and a scalar field source. The numerical results are 
supported by a qualitative dynamical analysis and a detailed calculation 
of the past asymptotic decay rates \cite{CHL}.

Finally, we note that this result corrects an earlier paper \cite{BD} 
in which two of the authors claimed the contrary to be true (i.e., brane\hs worlds
do not isotropize). The claim in \cite{BD} was based on the existence of the decaying 
mode $\propto a^{-3}$ (growing in the past) in the solutions for $\Delta$ 
and $Q^*$ which when substituted into the shear (\ref{eq:Sig}) and $div E$ 
constraints (\ref{eq:divE}) gave the same mode in the expansion normalised 
shear $\Sigma$ and electric part of the Weyl tensor ${\cal E}$. This resulted 
from using an incorrect propagation equation for the non\hs local energy flux  
in \cite{GM} (eqn. 32 in this paper) which gave different solutions for $Q^*$
\footnote{The authors of \cite{GM} agree with our findings; R. Maartens, private 
communication.}. The error in \cite{GM} results in a wrong decaying mode that 
is totally irrelevant for the future evolution of perturbations studied in 
that paper, but changes dramatically their behaviour as $t\rightarrow 0$. 
It is easy to see that when the corrected solutions (in Table 1) are 
substituted into (\ref{eq:Sig}) and (\ref{eq:divE}) the $a^{-3}$ mode actually 
cancels out in $\Sigma$ and ${\cal E}$. Note also that this mode does not
appear in the solution for the magnetic part of the Weyl tensor ${\cal H}$.

Since $\Sigma$, ${\cal E}$ and ${\cal H}$ give a better description 
of the geometry, their past evolution represents the 
true behaviour of anisotropies close to the initial singularity, 
and not $\Delta$. In fact one can argue that the existence of 
the $a^{-3}$ mode in $\Delta$ (defined in equation (\ref{eq:grads})) results 
from it not being the most appropriate measure of inhomogeneity at high 
energies since the dominant background energy density is not $\rho$ but 
$\rho^{tot}\propto\rho^2$ (see equation (11) in \cite{gmg}). More precisely, 
if we define $\Delta^{HE}$  (High Energy) normalising with $\rho^{tot}$ 
instead of $\rho$ then $\Delta^{HE}\propto\Delta/\rho$ and the decaying 
mode in $\Delta$ becomes a mode $\propto a^{-3 +3\gamma}$. 
Since $-3+3\gamma >0$ for $\gamma >1$, we remove the decaying mode 
in the density inhomogeneity. 

\acknowledgments
The authors would like to thank Yanjing He and W.C. Lim for useful discussions.
PKSD and NG thank the department of Mathematics and Statistics at Dalhousie 
University  for hospitality while some of this work was carried 
out and the NRF (South Africa) for financial support, NG thanks the 
University of Cape Town for a postgraduate scholarship and AAC acknowledges
NSERC (Canada) for financial support.
%%%%%%%%%%%%%%%%%%%%%%%%%%%%%%%%%%

\end{document}